\documentclass[pra,twocolumn,showpacs,preprintnumbers,amsmath,amssymb]{revtex4}
\usepackage{graphicx}
\usepackage{dcolumn}
\usepackage{bm}
\usepackage{latexsym,epsfig}

\newcommand{\beq}{\begin{equation}}
\newcommand{\eeq}{\end{equation}}
\newcommand{\bea}{\begin{eqnarray}}
\newcommand{\eea}{\end{eqnarray}}
\newcommand{\ben}{\begin{eqnarray*}}
\newcommand{\een}{\end{eqnarray*}}
\newcommand{\be}{\begin{enumerate}}
\newcommand{\ee}{\end{enumerate}}
\newcommand{\bfig}{\begin{figure}}
\newcommand{\efig}{\end{figure}}

\newcommand{\ba}{\begin{align}}
\newcommand{\ea}{\end{align}}

\newcommand{\D}{\displaystyle}
\newcommand{\la}{\langle}
\newcommand{\ra}{\rangle}
\newcommand{\aidag}{a_i^{\dagger}}
\newcommand{\nmin}{N_{\rm min}}
\newcommand{\nmax}{N_{\rm max}}
\newcommand{\Nmi}{N_{\rm MI}}
\newcommand{\Nsf}{N_{\rm SF}}

\newcommand{\vtrap}{V_{\rm t}}

\begin{document}

\preprint{}
\title{Signatures of the superfluid to Mott insulator transition in cold 
bosonic atoms in a one dimensional optical lattice}
\author{S.\ Ramanan}
\email{suna@cts.iisc.ernet.in} 
\affiliation{Centre for High Energy Physics,
Indian Institute of Science, Bangalore\ 560012, India}
\author{Tapan Mishra}
\email{tapan@iiap.res.in} \affiliation{ Indian Institute of
Astrophysics, II Block, Kormangala, Bangalore, 560 034, India.}
\author{Meetu Sethi Luthra}
\email{sethi.meetu@gmail.com} \altaffiliation [permanent address
]{Bhaskaracharya College of Applied Sciences, Phase-I,
Sector-2,Dwarka,Delhi,110075, India.}
 \affiliation{ Indian Institute of
Astrophysics, II Block, Kormangala, Bangalore, 560 034, India.}
\author{Ramesh V. Pai}
\email{rvpai@unigoa.ac.in} \affiliation{ Department of Physics, Goa
University, Taleigao Plateau, Goa 403 206, India. }
\author{B. P. Das}
\email{das@iiap.res.in} \affiliation{Indian Institute of
Astrophysics, II Block, Kormangala, Bangalore, 560 034, India.}

\date{\today}

\begin{abstract}
We study the Bose-Hubbard model using the finite size density matrix 
renormalization group method. We obtain for the first time a complete phase diagram for a 
system in the presence of a harmonic trap and compare it with that of the 
homogeneous system. 
To realize the transition from the superfluid to the Mott insulator phase we 
investigate different experimental signatures of these phases in quantities 
such as momentum distribution, visibility, condensate fraction and 
the total number of bosons at a particular density. The relationships between the various experimental signatures and the phase diagram are highlighted.
\end{abstract}

\pacs{03.75.Nt, 05.10.Cc, 05.30.Jp,73.43Nq}

\keywords{Suggested keywords}

\maketitle

\section{Introduction}
In recent years many theoretical and experimental investigations have been 
carried out in the field of ultracold atoms in optical 
lattices~\cite{dalfovo99,block08,lewenstein07}. 
The interest in bosonic systems began with 
the seminal paper by Fisher \textit{et al}~\cite{fisher}, 
where a phase transition 
from a superfluid (SF) to a Mott insulator (MI) in a lattice of bosons 
was predicted when the on-site Coulomb repulsion between the atoms dominates 
the nearest neighbor hopping amplitude. Since then, a variety of theoretical 
approaches~\cite{krauth,sheshadri,amico,kuhner,pandit,pai,kashurnikov,
batrouni,freericks,qmc1,qmc2} have been used to study the Bose-Hubbard Model 
(BHM)~\cite{fisher}.
There is good agreement between the phase diagrams obtained from the different techniques. 
While the BHM was originally developed in the context of $^4$He~\cite{fisher}, 
its potential to describe ultra cold bosons trapped in an optical lattice 
was soon realized by Jaksch \textit{et al}~\cite{jaksch}. 
This paper has had a great 
impact on the condensed matter community because high-precision 
experiments on cold atoms in traps can now  be used as a powerful and
reliable tool to study a variety of quantum phase 
transitions~\cite{dalfovo99,block08,lewenstein07}.
The experimental realization of the quantum phase transition from the 
superfluid 
to the Mott insulator in three dimensions~\cite{greiner}, 
two dimensions~\cite{spielman}, as well as in one
dimension~\cite{stoferle} soon followed. The bosons in an optical lattice
are well described by the Bose-Hubbard model modified to include a 
trap potential~\cite{jaksch}, which is normally harmonic. 
In the presence of a trap, the density profile exhibits a rich structure as the SF and the MI phases coexist. A variety of numerical methods have been applied 
to understand the 
model~\cite{wessel,svistunov,stoof,bergkvist,pollet,smita,mitra}. 
The most important aspect that has emerged is the lack of a global phase in 
these systems. 
As an analogy, in three dimensions the superfluid and the Mott insulator phases 
coexist as shells in an onion.
The unprecedented control over the system parameters by tuning the laser 
intensity has paved the path for the experimental realization of these 
predictions~\cite{bloch06,campbel}.

In this paper, we re-visit the one dimensional Bose-Hubbard model using
 the density matrix renormalization group method. Our main motivation is to 
obtain the phase diagram, given the experimental realization of the shell 
structure and their signatures.
The Bose-Hubbard model, describing bosons in an optical lattice, is given by 
\bea
\label{eq:ham}
H&=&-t\sum_{<i,j>}(a_{i}^{\dagger}a_{j}+h.c)\nonumber\\ 
&& \mbox{} +\frac{U}{2}\sum_{i}
n_{i}(n_{i}-1)+\vtrap\sum_{i}r_i^2n_i.
\eea
where $t$ is the hopping amplitude, $U$ is the on-site repulsion between 
the atoms and $\vtrap$ is the depth of the external trapping potential, 
$a_i^\dagger(a_i)$ are the bosonic creation (annihilation) operator, 
$n_i=a_i^\dagger a_i$ is the number operator and $r_i$ is the position of the 
$i^{\rm th}$ lattice site from the trap center. For simplicity, the energy is scaled in 
units of the hopping amplitude which is set to one. 

In experiments, the optical lattice potential ($V_{\rm OL}$), formed by the superposition of three 
counter propagating laser beams~\cite{block08, gerbier_vis}, can be written as:
\beq
	V_{\rm OL} = V_0 (\sin^2(k_Lx) + \sin^2(k_L y) + \sin^2(k_Lz))
	\label{op_lat_eqn}
\eeq
where $V_0$ is the lattice depth measured in units of the single photon recoil energy
 $E_R$, $k_L = 2 \pi/\lambda_L$ is the wave vector and $\lambda_L$ is the laser wavelength. 
By increasing the intensity of the beam in two directions, the hopping is
 restricted to one dimension and this results in a one dimensional lattice.
 The on-site interaction $U$ and the hopping amplitude $t$ are related to 
$V_0$ and $E_R$ as follows:
\beq
	\D\frac{U}{t} \propto \exp(2 V_0/E_R)^{1/2}.
	\label{op_lat_eqn1}
\eeq
Therefore, by varying the lattice depth the ratio of $U/t$ can be tuned.

Earlier studies of the
homogeneous system show that when the on-site interaction strength $U$ is 
small compared to the hopping amplitude $t$, the system remains in the SF 
phase characterized by long range coherence. When $U$ increases and 
becomes much larger than $t$, a transition from the SF to the MI phase occurs at 
some critical value of  $U = U_c\sim3.4$ (in units of $t$)~\cite{kuhner,pandit,pai}. 
This transition belongs to the Kosterlitz-Thouless universality 
class~\cite{kt,giamarchi}. 
The SF-MI transition gets more interesting in the presence of a 
trap~\cite{qmcwork,schollwock1,epl,hrk}. 
In this case, 
the entire system remains in the SF phase for small $U$ values, 
but as $U$ increases, a MI phase develops around a central SF phase, 
followed by a SF shoulder. Further increase in $U$ ultimately results in a 
MI phase thoughout the lattice 
with the exception of the trap edges, which we refer to as a central MI phase. This alternating occurrence of the SF 
and the MI phases can be observed in the number density profile. 
The system exhibits a plateau at integer densities (MI phase) surrounded by a 
region of non-integer densities (SF phase)~\cite{wessel,svistunov,stoof,bergkvist,pollet,smita,mitra,qmcwork,schollwock1,epl,hrk} and this has
been recently observed in experiments~\cite{bloch06,campbel} 
in 3D optical lattices. 

In this paper, we first obtain the phase 
diagram of the model, given by eqn.~(\ref{eq:ham}), in the 
homogeneous ($\vtrap=0$) limit.  
We then extend our analysis to the inhomogeneous
case and compare it with the homogeneous case. Finally we obtain experimentally measurable quantities like
condensate fraction, visibility and density profile and deduce the inhomogeneous  phase diagram.
We also discuss how the
shell structure in the optical lattice can be observed from condensate fraction and visibility.

This paper is structured as follows: Section~(\ref{sec2}) contains a brief 
discussion on the Finite Size Density Matrix Renormalization Group technique 
(FS-DMRG). Section~(\ref{sec3}) discusses the phase diagram obtained from the 
number density profile, while Section~(\ref{sec4}) analyzes the experimental 
signature for the phase transition and we summarize our results in Section~(\ref{sec5}). 

\section{Method of calculation}
\label{sec2}

We have employed the finite size density matrix renormalization group method (FS-DMRG)~\cite{white,schollwock} with open-boundary 
condition to determine the 
ground state energy and the wave function of the system. 
This method is one of the most powerful 
technique in one dimension and has been widely used to study the 
Bose-Hubbard model~\cite{pandit,pai,kuhner,schollwock1,epl}.
We have considered a soft-core case by retaining four bosonic states per site and 
the weight of the states neglected in the density matrix formed from the 
left or right blocks is less than $10^{-6}$. For better convergence of the 
ground state energies of various phases, we have performed a 
finite size sweeping procedure~\cite{white,pandit}, twice in each iteration 
of the FS-DMRG method. 

Our FS-DMRG method consists of two steps: (i) a DMRG iteration where the length
of the system $L$ is increased to $L+2$ and (ii) finite size sweeping
to achieve better convergence of the ground state energy 
$E_L(N)$~\cite{white,pandit}.   
We consider a system with an initial length $L=4$ and the number of bosons 
$N=4$. The FS-DMRG
procedure is then employed, keeping the density of bosons fixed at $\rho=1$, 
until we have a desired number of bosons in the system, say $N=30$. 
Then onwards, in the FS-DMRG iteration, only the 
length of the system 
is increased keeping $N$ fixed, until the system grows to a desired size, 
for example $L=100$. At this point we have a system of length
$L=100$ with $N=30$ bosons.
Now keeping the length fixed,
we increase $N$ in steps of $1$ at the end of each FS-DMRG sweep.
In our example, $N$ is varied from $30$ to $150$. Thus at the end of 
the FS-DMRG calculation, 
for a given set of parameter values,  
we have ground state energies $E_L(N)$ and wave functions $|\psi_{LN}\ra $
for a system of length $L=100$ 
with $N$ varying from $30$ to $150$.  
From $E_L(N)$, we obtain the chemical potential $\mu$ of the 
system 
 \beq
 \mu = \D\frac{\delta E_L(N)}{\delta N}.
\label{eq:mu}	 
\eeq
The on-site local number density $\la n_i \ra$ and the local compressibility $\kappa_i$ are defined as 
\beq
	\la n_i \ra= \langle\psi_{LN}|n_i|\psi_{LN} \rangle 
\eeq 
\beq
	\kappa_i =\D\frac{\delta n_i}{\delta \mu}.
\label{eq:kappa}
\eeq

For the homogeneous case, the local density $\langle n_i \rangle=N/L$ and hence the compressibility 
are uniform throughout the system except at the edges due to boundary effects. 
The chemical potential corresponds to that of the whole
system. However, for finite $\vtrap$ the lattice is inhomogeneous 
and as a result the
local density $\langle n_i \rangle$ attains its maximum value at the center of the trap
 where the potential is minimum and decreases as we move away from the center, eventually going to zero. 

In most of our calculations, we have taken $\vtrap=0.004$ and $0.008$, $L=200$, with  
$N$ ranging between
$30$ and $150$ and $U$ between $2$ and $20$. 

\section{Results and Discussions} 
\label{sec3}
\subsection{Homogeneous case}

\bfig[htbp]
\begin{center}
	\includegraphics[width = 3.4in, angle = 0, clip = true]
{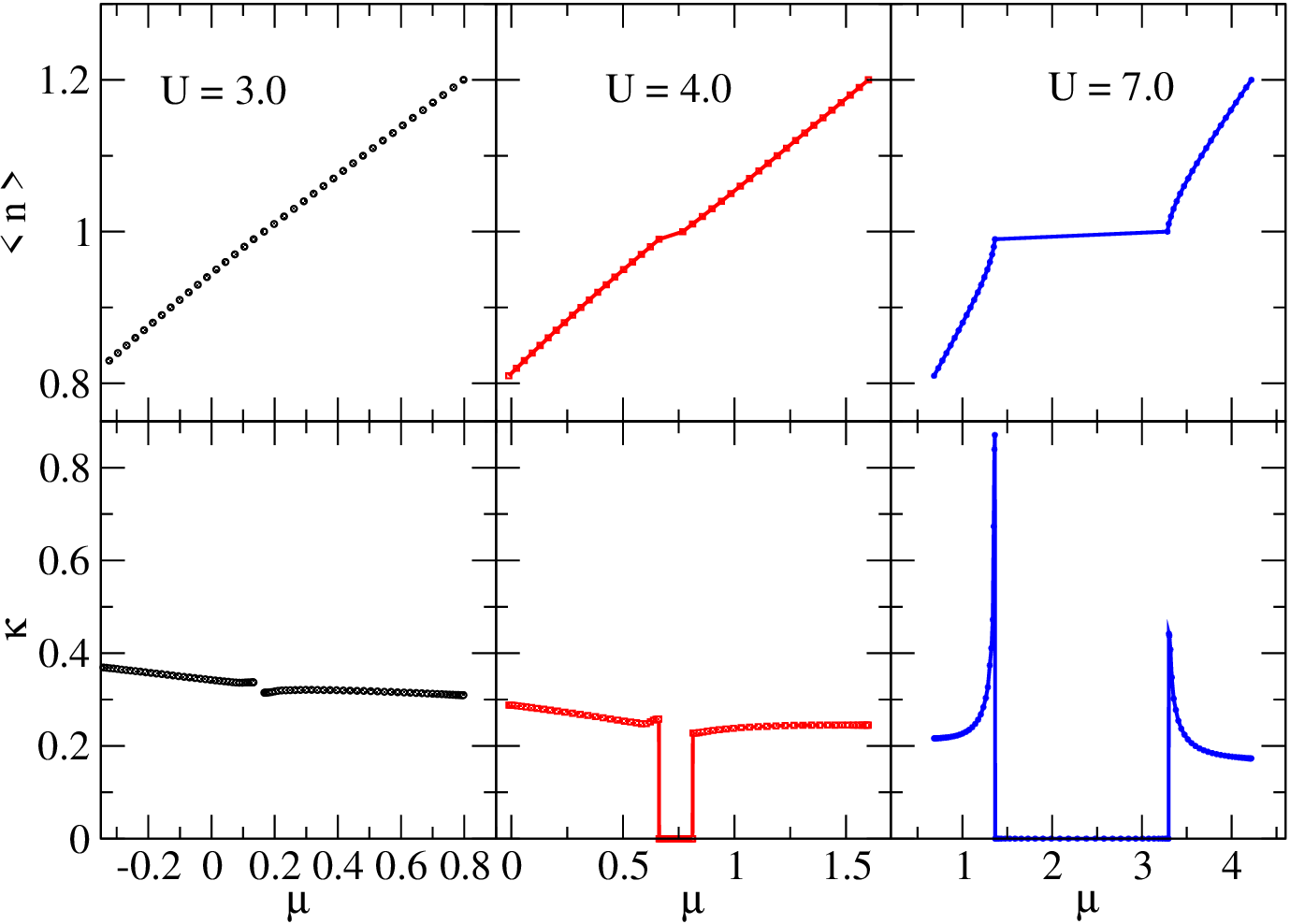}
	\caption{Variation of the compressibility $\kappa$ (lower panels) 
and the density $\langle n \rangle$ (upper panels) as a function of $\mu$ 
for the homogeneous case ($\vtrap=0$). The plateau region at 
$\langle n \rangle=1$ and $\kappa=0$ signals the incompressible ($\rho =1$) MI phase.
}
	\label{fig:fig1}
\end{center}
\efig

\bfig[htbp]
\begin{center}
	\includegraphics[width = 3.4in, angle = 0, clip = true]
{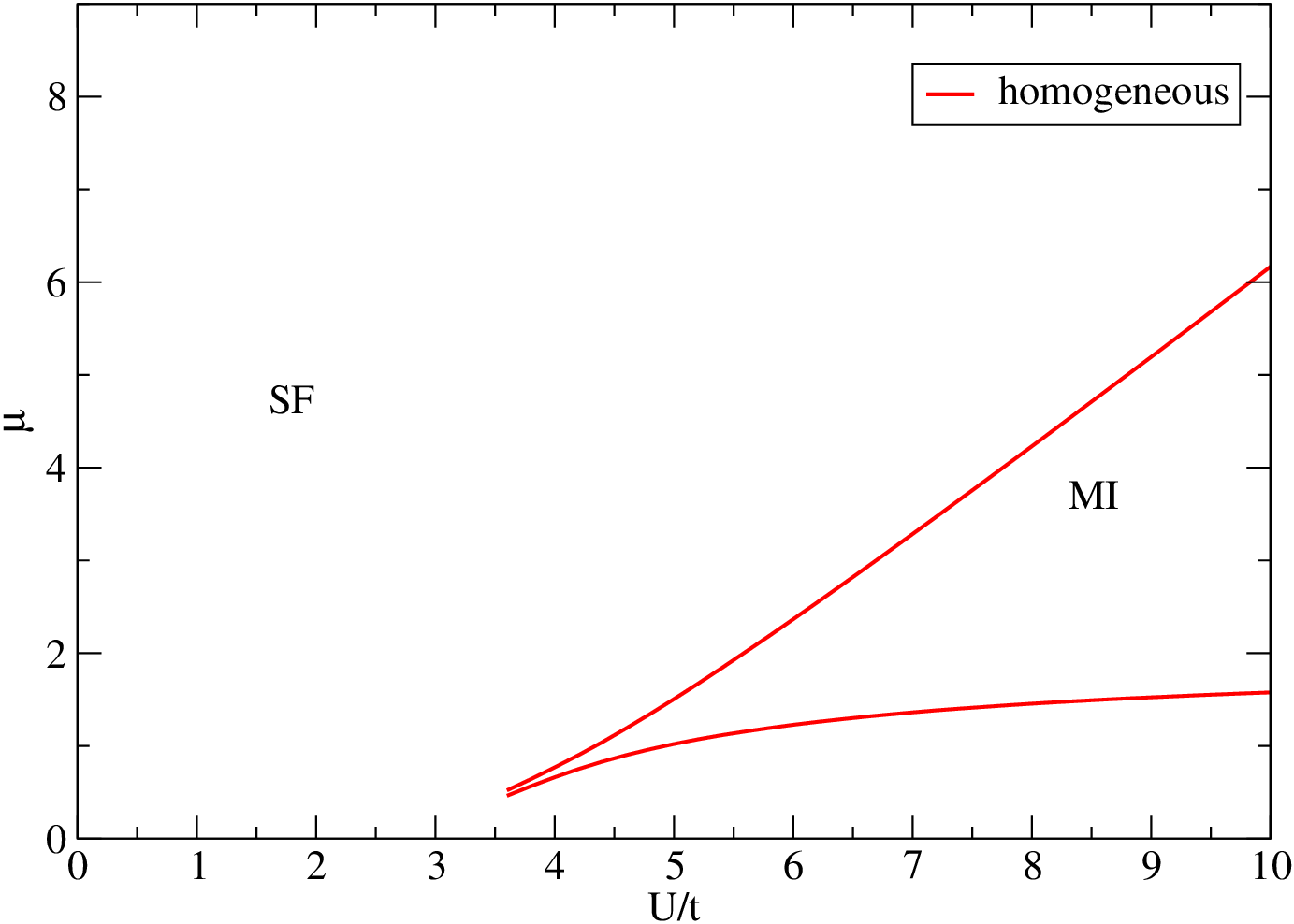}  
	\caption{Phase diagram for the homogeneous Bose-Hubbard model. The MI phase has density $\rho=1$.}
     	\label{fig:fig2}
\end{center}	
\efig
We begin our discussions with the homogeneous case. The density of the system $\la n \ra$ as a function of the chemical potential $\mu$ for  
three values of $U=3,4$ and 
$7$ are shown in the top panels of Fig.~(\ref{fig:fig1}). The formation of a 
plateau in the $\la n \ra$ versus $\mu$ plots, for $U=4$ and $7$ at $\langle n\rangle =1$, in contrast to $U=3$,
signals the onset of the $\rho=1$
MI phase, where $\rho$ denotes the density per site. The lower panels in Fig.~(\ref{fig:fig1})
show the compressibility $\kappa$ (calculated using Eq.~(\ref{eq:kappa})) as a function of $\mu$. 
It is clear that the MI phase is incompressible, i.e., $\kappa = 0$. The cusp in $\kappa$ as $\langle n \rangle $ approaches 
$1$, shown in 
Fig.~(\ref{fig:fig1}), is due to quantum criticality.
The phase diagram for the homogeneous system is obtained by picking out the 
values of $\mu$ at the knees where $\langle n \rangle=1 $ and $\kappa =0$
and plotting them in the $\mu - U $ plane. 
This is shown in the fig.~(\ref{fig:fig2}) and is in agreement with earlier
results~\cite{pandit,pai,kuhner,freericks}.
The cusp in the compressibility reflects the Kosterlitz-Thouless type behavior of the SF-MI transition.

\subsection{Inhomogeneous case}

\begin{figure}[htbp]
\begin{center}
	\includegraphics[width = 3.4in, angle = 0, clip = true]
{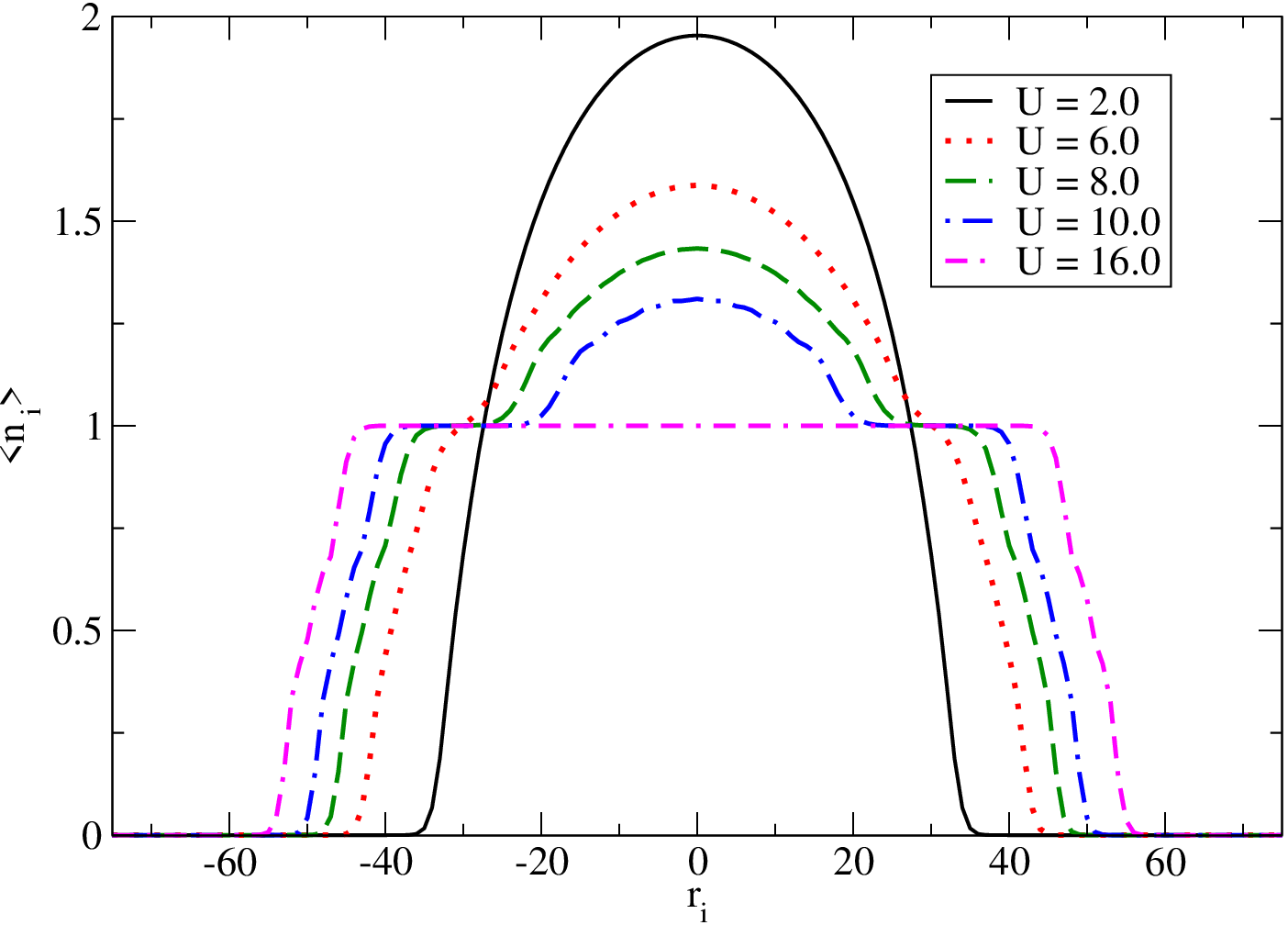}
  	\caption{Density profile as a function of $r_i$ for 
$\vtrap = 0.004$, $L = 200$ and $N = 100$ for a range of $U$. Note that as the on-site repulsion increases, a MI phase forms around the central SF phase, finally leading to a central MI phase for higher values of $U$.}
     	\label{fig:fig3}
\end{center}
\end{figure}

\begin{figure}[htbp]
\begin{center}
	\includegraphics[width = 3.4in, angle = 0, clip = true]
{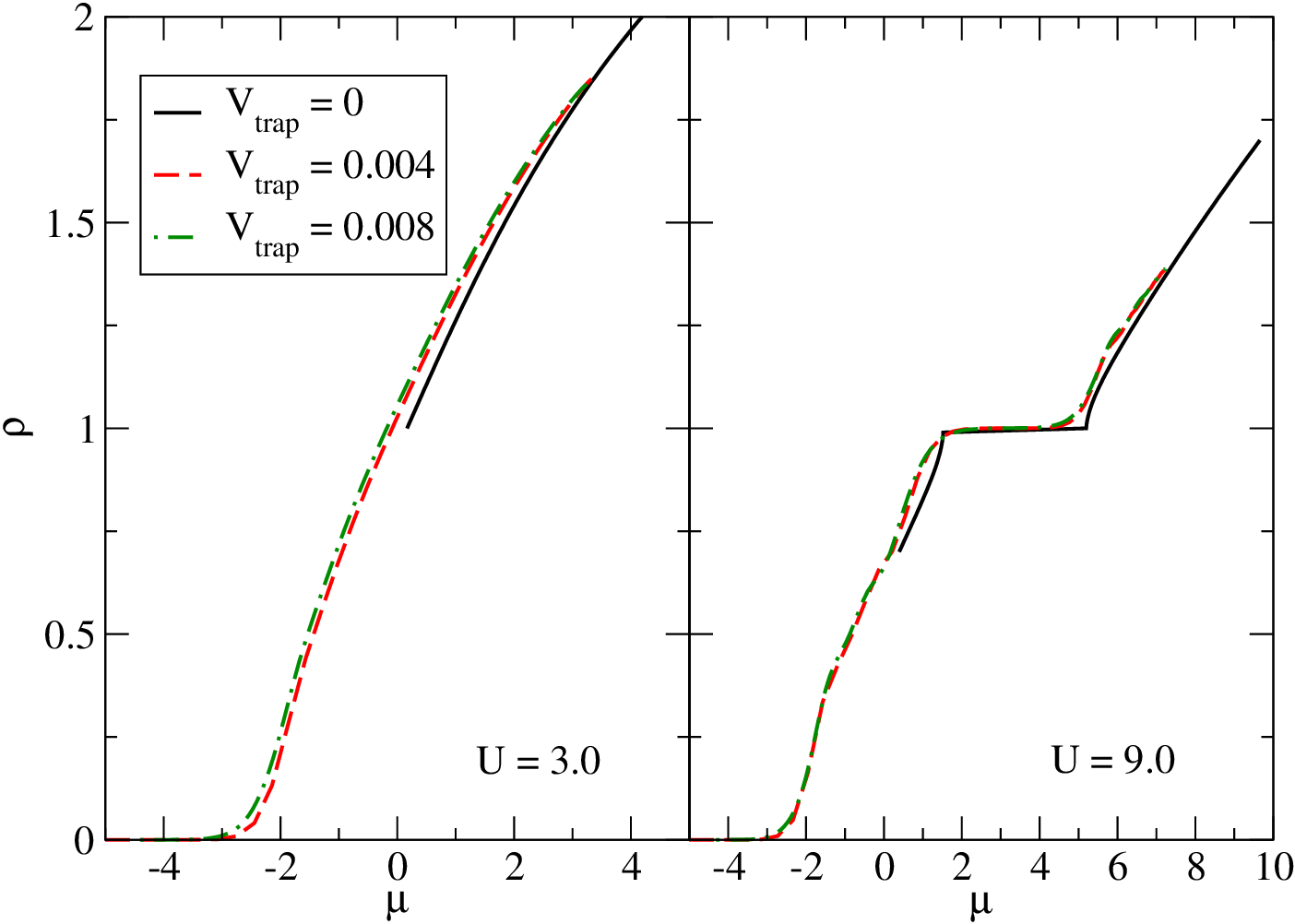}
  	\caption{Density as a function of chemical 
potential for the homogeneous and inhomogeneous cases. The solid line represents the homogeneous system, the dashed lines and dashed-dot lines, the inhomogeneous case ($\vtrap = 0.004$ and $\vtrap = 0.008$ respectively). The 
homogeneous case has sharp transitions compared to the finite trap case.}
     	\label{fig:fig3b}
\end{center}
\end{figure}

We now analyze the case when there is a finite trapping potential $\vtrap$ in the Bose-Hubbard Hamiltonian. 
Taking the depth of the trap $\vtrap=0.004$, number of bosons $N=100$ and length $L=200$,
we obtain the local density profile $\langle n_i \rangle$ as a function of the distance from the center of the trap $r_i$, as shown in Fig.~(\ref{fig:fig3}). In contrast to the homogeneous case, 
$\langle n_i \rangle$ is not
uniform when the trap is finite. It decreases 
monotonically as we move from the center of the trap towards the edges. For 
larger values of $U$ ($U > 6$) the density profile  develops a well defined 
plateau at $\langle n_i \rangle=1$ and the length of the plateau grows as 
$U$ is increased further. 
A simple way to understand the 
behavior of  $\langle n_i \rangle$ is through the Local Density Approximation
(LDA)~\cite{bergkvist,hrk} where the local density at site $i$ in the trapped case
is given by the density of a homogeneous system with a chemical potential 
\beq
	\mu_i = \mu - \vtrap r_i^2.
	\label{eqn:mu_loc}
\eeq 
Here $\mu$ refers to the chemical potential at the center of the lattice where
the trap potential is zero.
Using Eq.~(\ref{eqn:mu_loc}) we can re-scale the x-axis of Fig.~(\ref{fig:fig3}), so that we get the local density as a function of the local chemical potential, as shown in Fig.~(\ref{fig:fig3b}) for $U=3$ and $9$, where the homogeneous case is also documented for comparison.
The local density 
as a function of the local chemical potential exhibits trends similar to the 
homogeneous system, that is the value of $\la n_i \ra$ increases smoothly as
$\mu_i$ increases for 
lower values of $U$, where the entire system is in the compressible 
superfluid phase. 
However, as $U$ increases further, a clear plateau emerges at 
$\la n_i \ra = 1$ indicating the onset of the $\rho=1$ MI phase.

\begin{figure}[htbp]
  \begin{center}
  \includegraphics[width = 3.4in, angle = 0, clip = true]
{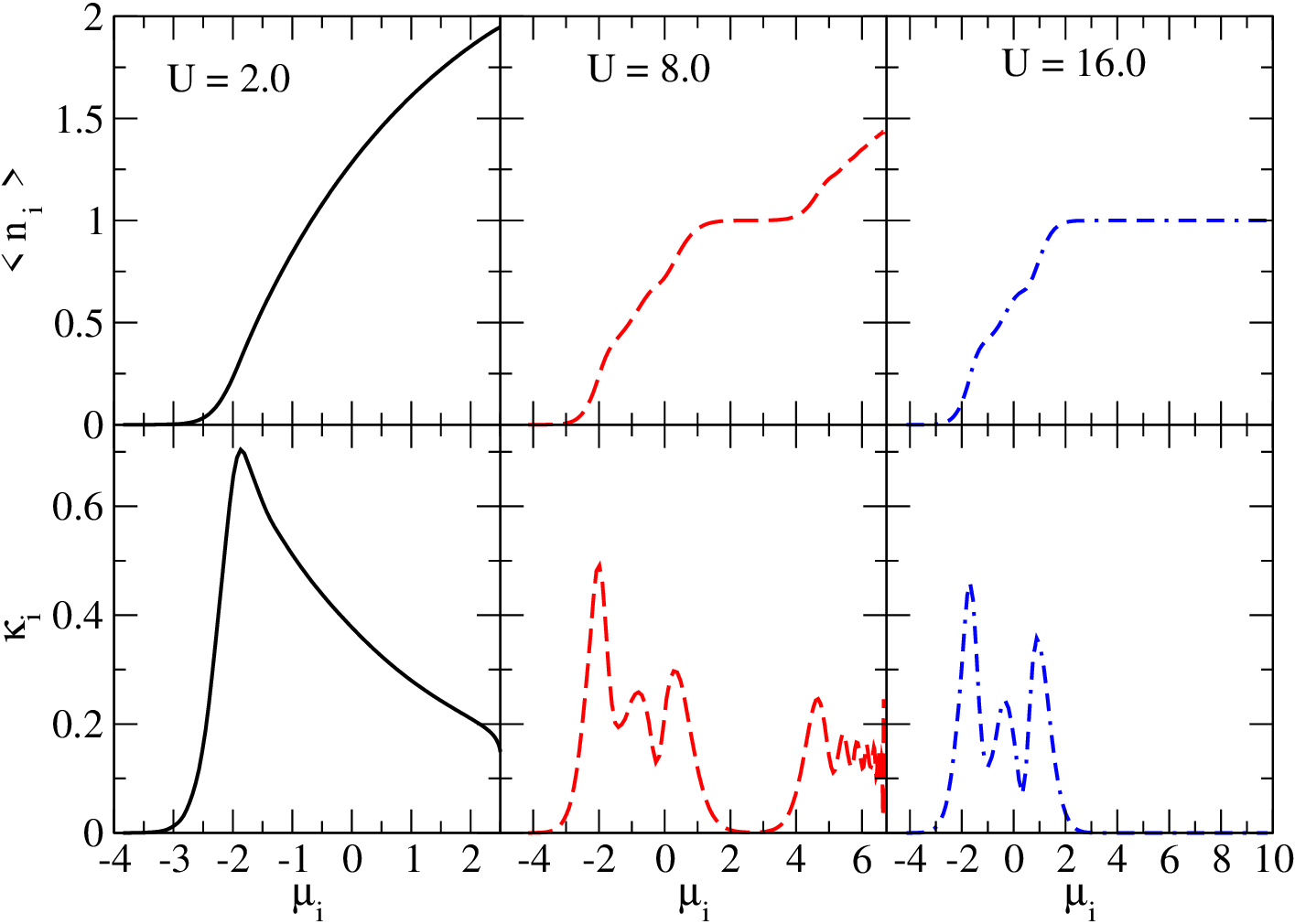}
  \caption{Local density and local compressibility as a function of the 
local chemical potential for $\vtrap=0.004$.}
     \label{fig:fig4}
\end{center}
\end{figure}

The local compressibility can be obtained from the local density using eqn.~(\ref{eq:kappa}). Fig.~(\ref{fig:fig4}) shows the local density and the local compressibility in the upper and the lower panels, respectively.
The plateau region at $\langle n_i \rangle =1$  
(upper panels) has the corresponding 
local compressibility equal to zero. 
This verifies that the plateau present in the number density profile represents the MI phase. Furthermore, this confirms the coexistence of 
the SF and the MI phases in the presence of a trap. Though there is an overall agreement between  
the homogeneous and the inhomogeneous cases, we note that there are slight discrepancies.
The sharp SF-MI transition observed in the homogeneous system is 
smoothened out in the presence of a trap.
The cusp like behavior observed in $\kappa$ for the homogeneous case is also
lost. 
The agreement between the homogeneous and the inhomogeneous cases 
prompted us to obtain the phase diagram for the inhomogeneous system, by analogously picking out the 
values of $\mu$ at the knees where $\la n_i \ra = 1$, 
and plotting them in the $\mu - U $ plane.
The resultant phase diagram is compared with the homogeneous result in Fig.~(\ref{fig:fig5}).  
It is interesting to note that the MI lobe 
for the inhomogeneous system lies within that of the homogeneous system. 
However, the MI lobe retains its cusp like shape. 
Thus the SF-MI transition for
density $\rho=1$ and $\vtrap >0$ does have the Kosterlitz-Thouless universality class behavior.

\bfig[htbp]
  \begin{center}
  \includegraphics[width = 3.4in, angle = 0, clip = true]{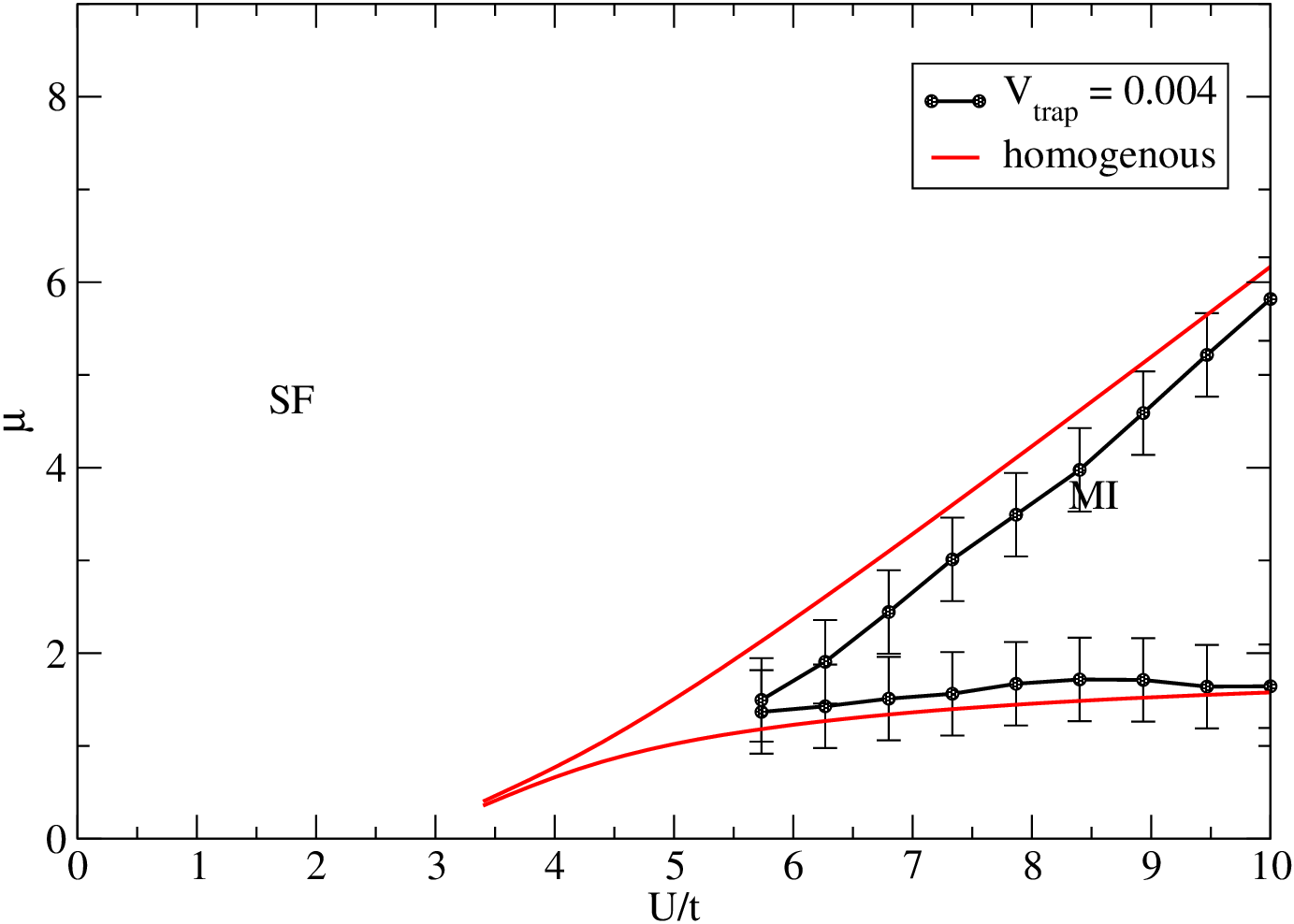}
  \caption{Homogeneous and inhomogeneous phase diagram}
     \label{fig:fig5}
\end{center}	
\efig

\section{Experimental Signatures}
\label{sec4}

In the earlier section we have established that the ground state 
phases for the Bose-Hubbard Hamiltonian given by eqn.~(\ref{eq:ham}) are either the 
superfluid or the Mott insulator depending on the ratio of 
$U/t$ and the local chemical potential $\mu_i$. Since $\mu_i$ is uniform for 
the homogeneous system, the ground state is global in nature. However, 
for inhomogeneous systems, $\mu_i$ being non uniform, both the SF and the MI 
phases coexist, as already discussed. It would be worthwhile to explore the signatures of these coexisting phases in experimentally determined quantities. 

It is now possible in experiments to record the spatial distribution of the lattice with different 
filling factors using spatially selective 
microwave transitions and spin-changing collisions as shown by 
F\"{o}lling et al.~\cite{bloch06}. Similar experiments in one-dimensional optical 
lattices can yield density profiles from which the phase diagram can be obtained, as discussed in the previous section. 
Another way to obtain direct information about the Mott plateaus (shells in 3D)
is through the atomic clock shift experiment~\cite{campbel}. 
By using density dependent 
transition frequency shifts, sites with different occupation can be
spectroscopically distinguished, thus giving us the information about the number of 
sites with a given density $\rho$, defined as $N(\rho)$. 
In  Fig.~(\ref{fig:fig7})
we plot $N(\rho)$ versus $\rho$ for several values of $U$ for a system with 
$\vtrap=0.004$ and $N=100$. The peak in 
$N(\rho)$ at $\rho=1$ for $U>6.0$ is a direct signature of a well developed
Mott insulator plateau in the inhomogeneous 
system. The size of this  peak 
increases with $U$ and is consistent with the increase in the length of the MI plateau
as shown in the phase diagram (see Fig.~(\ref{fig:fig5})). From 
$N(\rho)$ we obtain the following quantities: $\Nmi/N$, the fraction of 
number of bosons in the Mott plateau and  $\Nsf/N$, the fraction
of bosons in the SF region. Here $\Nmi$ and $\Nsf$ are the 
number of bosons in the MI and the SF phases respectively. Fig.~(\ref{fig:fig8}) 
shows both $\Nmi/N$ and $\Nsf/N$ for 
several values of $U$ keeping $\vtrap=0.004$. For $U < 6.0$,
we observe that $\Nsf/N$ is close to one, while $\Nmi/N$ is close to zero. This 
is because the entire system is in the SF phase. 
However, for $U > 6.0$, the increase in $\Nmi/N$, signals the formation of a MI plateau in the system. 
The critical value of $U$, marking the transition to the MI phase ($\rho=1$) can be read-off from Fig.~(\ref{fig:fig8}) and is given by $U_C\sim 6$. The small plateaus seen in $\Nmi/N$ and $\Nsf/N$ are indicative of the detailed distribution of the bosons as $U$ increases in the presence of a trap. 

\bfig[ht]
\begin{center}
\includegraphics[width = 3.4in, angle = 0, clip = true]{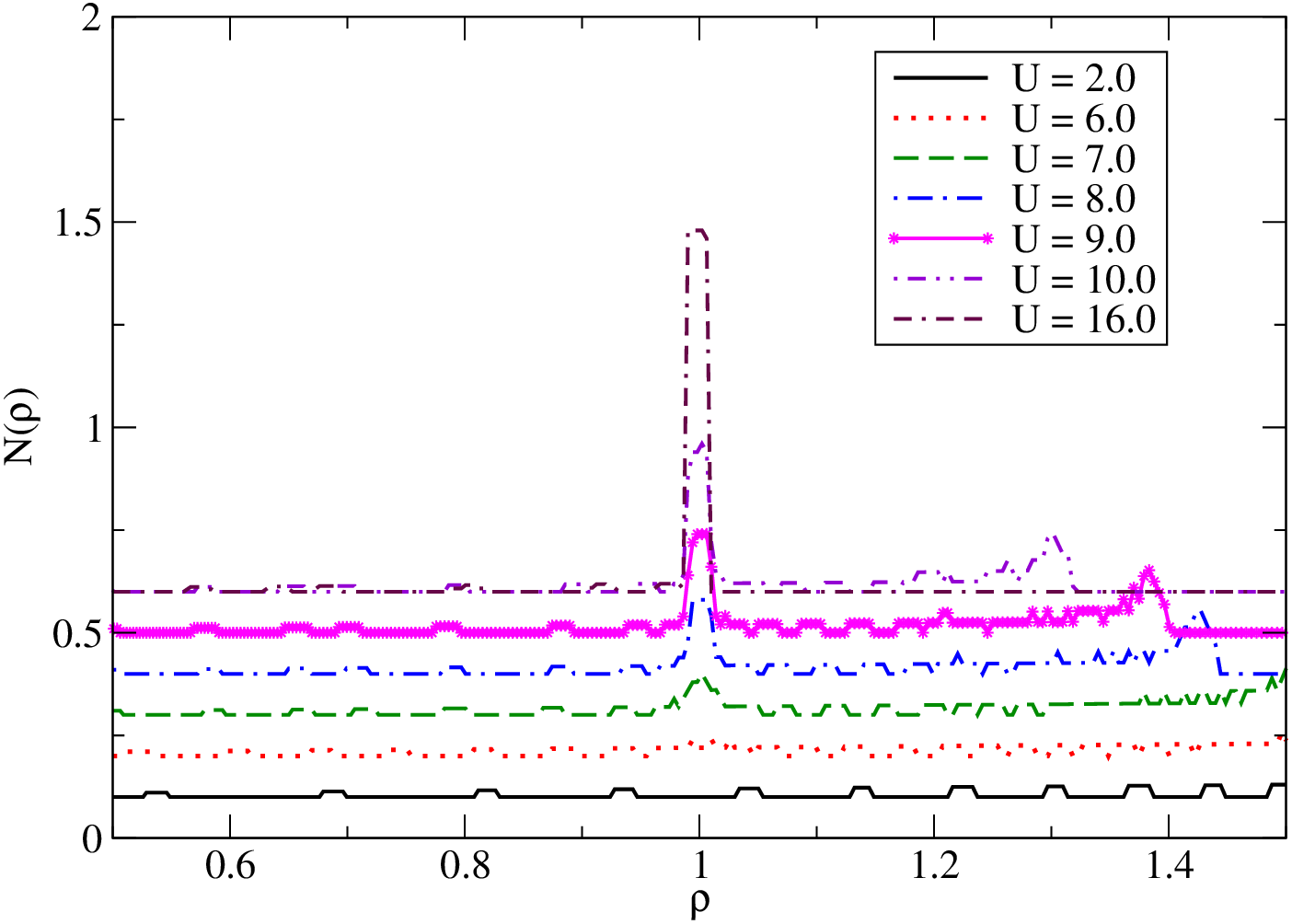}
\caption{$N(\rho)$ versus $\rho$ for $\vtrap=0.004$. A small offset is added to the Y-axis for clarity. The peak in $N(\rho)$ at $\rho=1$ signals the MI plateau in 
density profile (Fig.~\ref{fig:fig3}) and thus confirms the coexistence of SF 
and MI phases.}
\label{fig:fig7} 
\end{center}
\efig

\bfig[ht]
\begin{center}
\includegraphics[width = 3.4in, angle = 0, clip = true]{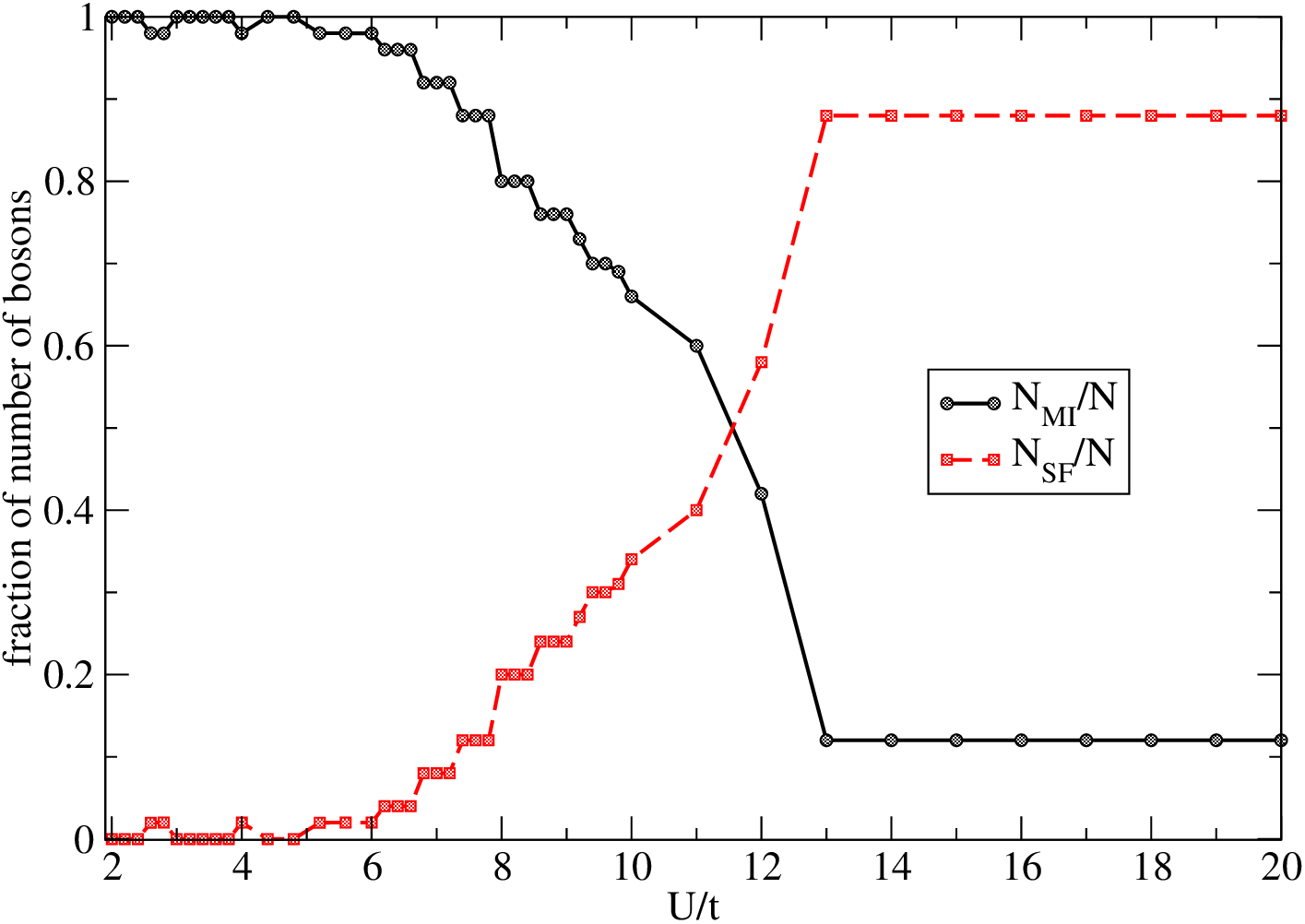}
\caption{$\Nmi/N$ and $\Nsf/N$ as function of $U$. The critical on-site
interaction $U_C \sim 6.0$ for the SF-MI transition for $\rho=1$ can be easily read-off.}
\label{fig:fig8} 
\end{center}
\efig
 
In other experiments, the cold atom gas
trapped in 
an optical lattice is allowed to expand and the interference pattern in the 
density of the expanding gas is recorded. The density 
distribution is mirrored in the momentum distribution defined as,
\beq
	n(q) = \D\frac{1}{L} \sum_{k, l = 1}^L \la 
a_k^{\dagger} a_l \ra \exp(i q (k - l)) 		
	\label{eqn:mom_dist1}
\eeq
where $k,l$ are the lattice sites. Fig.~(\ref{fig:fig9}) shows the momentum distribution for different $U$ values. The superfluid phase that has 
long-range coherence exhibits sharp interference peaks, while the Mott 
insulator phase, where the local density is pinned to integer values per site, 
breaks this coherence and hence no sharp peaks are observed~\cite{greiner}. 
Although the presence of the interference peaks in the density distribution 
(or analogously momentum distribution) was originally used to signal the 
formation of a SF phase, recently it has been established that the 
visibility of the interference fringes~\cite{nandini, ho, sengupta, gerbier_vis} 
provides a clear signature of the transition. The fringe visibility is defined as 
\beq
	{\cal{V}} = \D\frac{\nmax - \nmin}{\nmax + \nmin}
	\label{eq:vis1}
\eeq
where $\nmax$ and $\nmin$ are the maximum and the minimum of the momentum 
distribution measured at $q = \pm 2 \pi$ and $q = \pm \pi$ respectively, in one dimension.
The condensate fraction, that is the number of bosons in the condensate with respect to the total number of bosons, is defined as the largest eigenvalue of the matrix $\la \aidag a_j \ra$ divided by the total number of bosons~\cite{leggett_rev}. The fringe visibility and the corresponding condensate fraction for $\vtrap=0.004$, $N=100$ and $\vtrap=0.008$, $N=50$
are given in 
Figs.~(\ref{fig:fig10}) and (\ref{fig:fig11}), respectively.

\bfig[ht]
\begin{center}
 \includegraphics[width = 3.4in, angle = 0, 
clip = true]{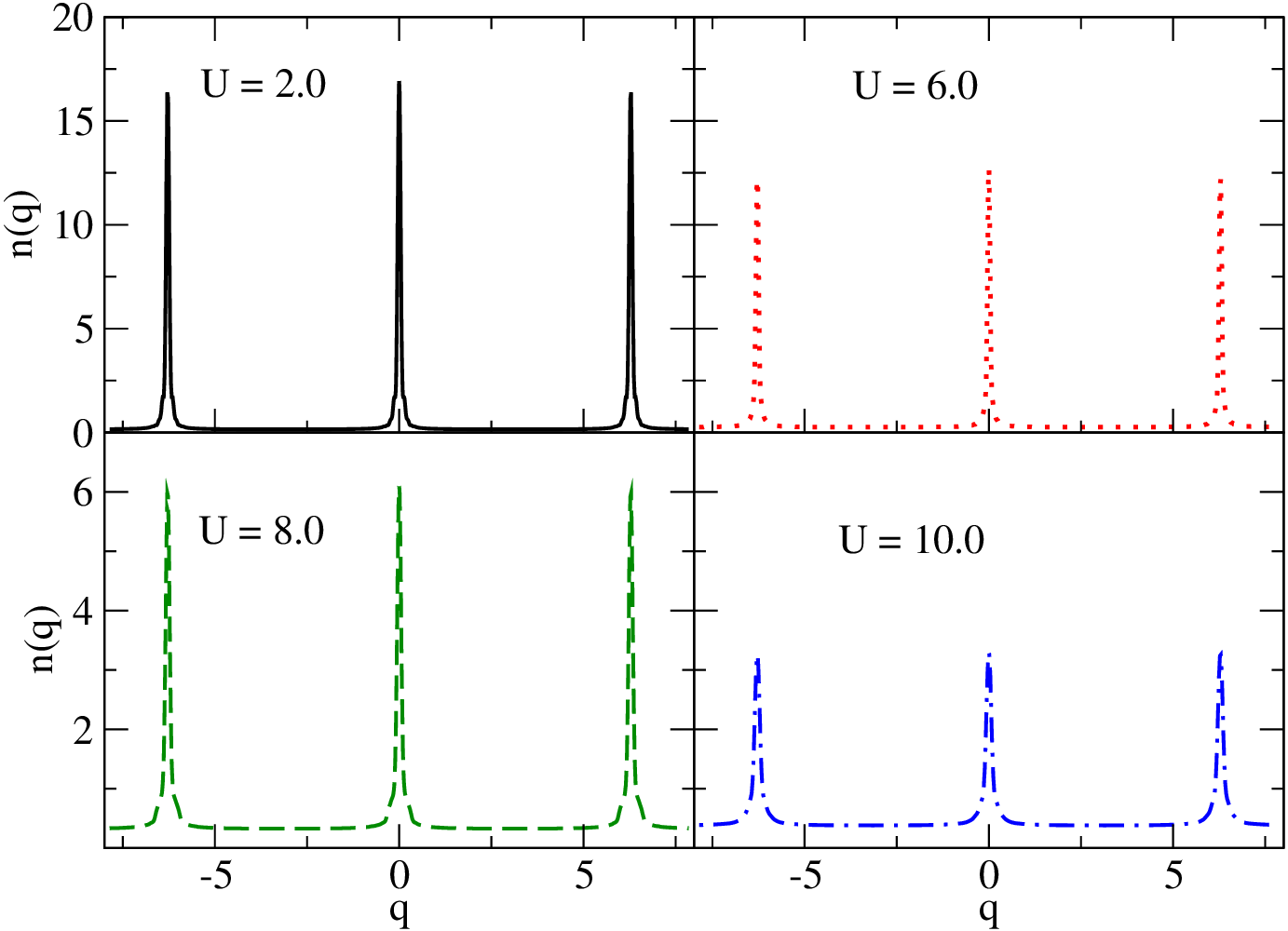}
 \caption{Momentum distribution as a function of $q$ in units of the lattice 
spacing. Note that at smaller $U$ values there are three prominent interference peaks at $q = 0$ and 
$q = \pm 2 \pi$. As $U$ increases, the peaks get smaller, indicating a 
transition from SF to MI.}
\label{fig:fig9}
\end{center}
\efig

\bfig[ht]
\begin{center}
 \includegraphics[width = 3.4in, angle = 0, 
clip = true]{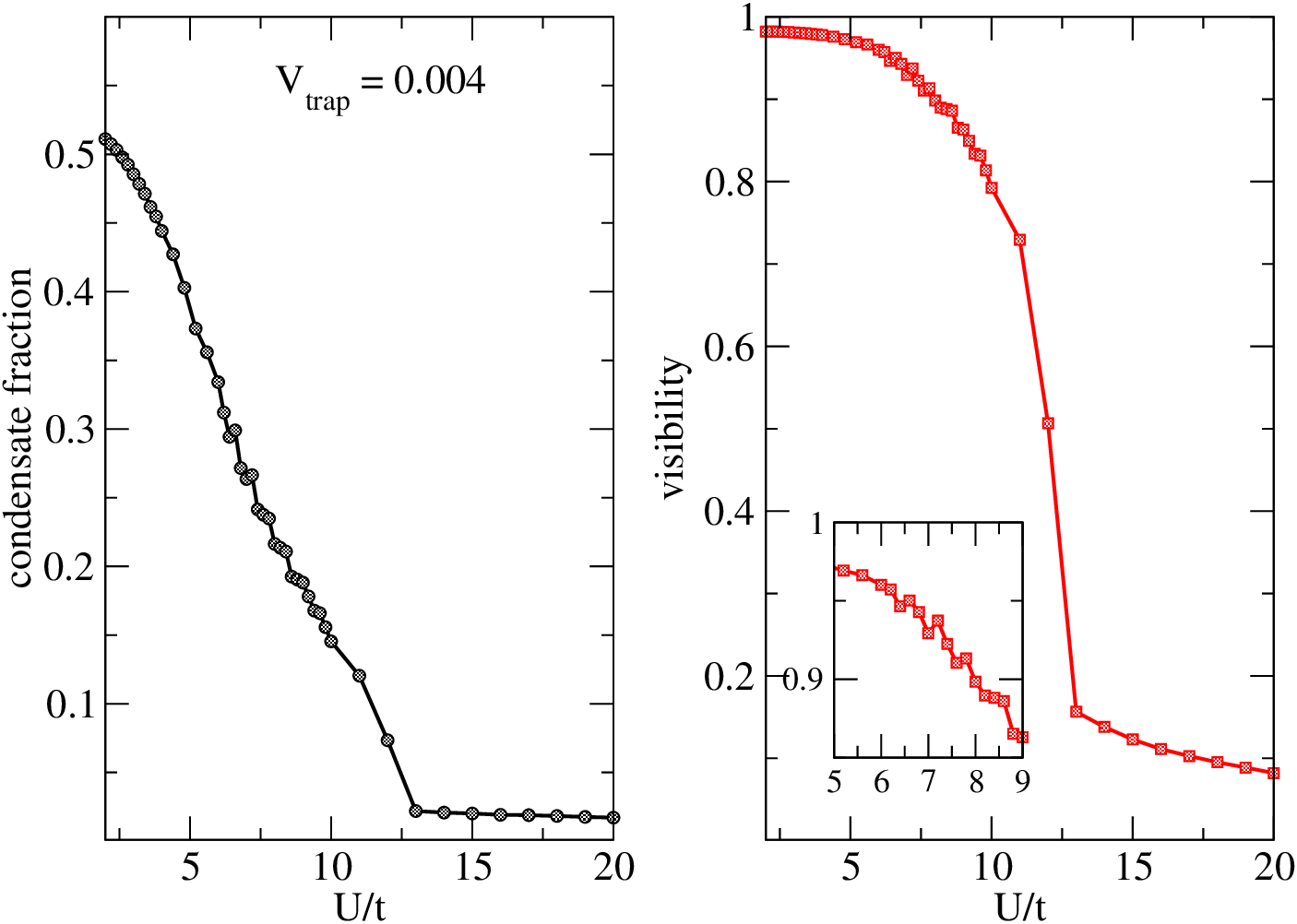}
 \caption{Condensate fraction (left) and visibility (right) as a function of $U$ for a trap of depth $\vtrap = 0.004$. The inset zooms in on the kinks observed in the visibility corresponding to the formation of Mott shoulders.}
 \label{fig:fig10}
\end{center}
\efig

\bfig[ht]
\begin{center}
 \includegraphics[width = 3.4in, angle = 0, 
clip = true]{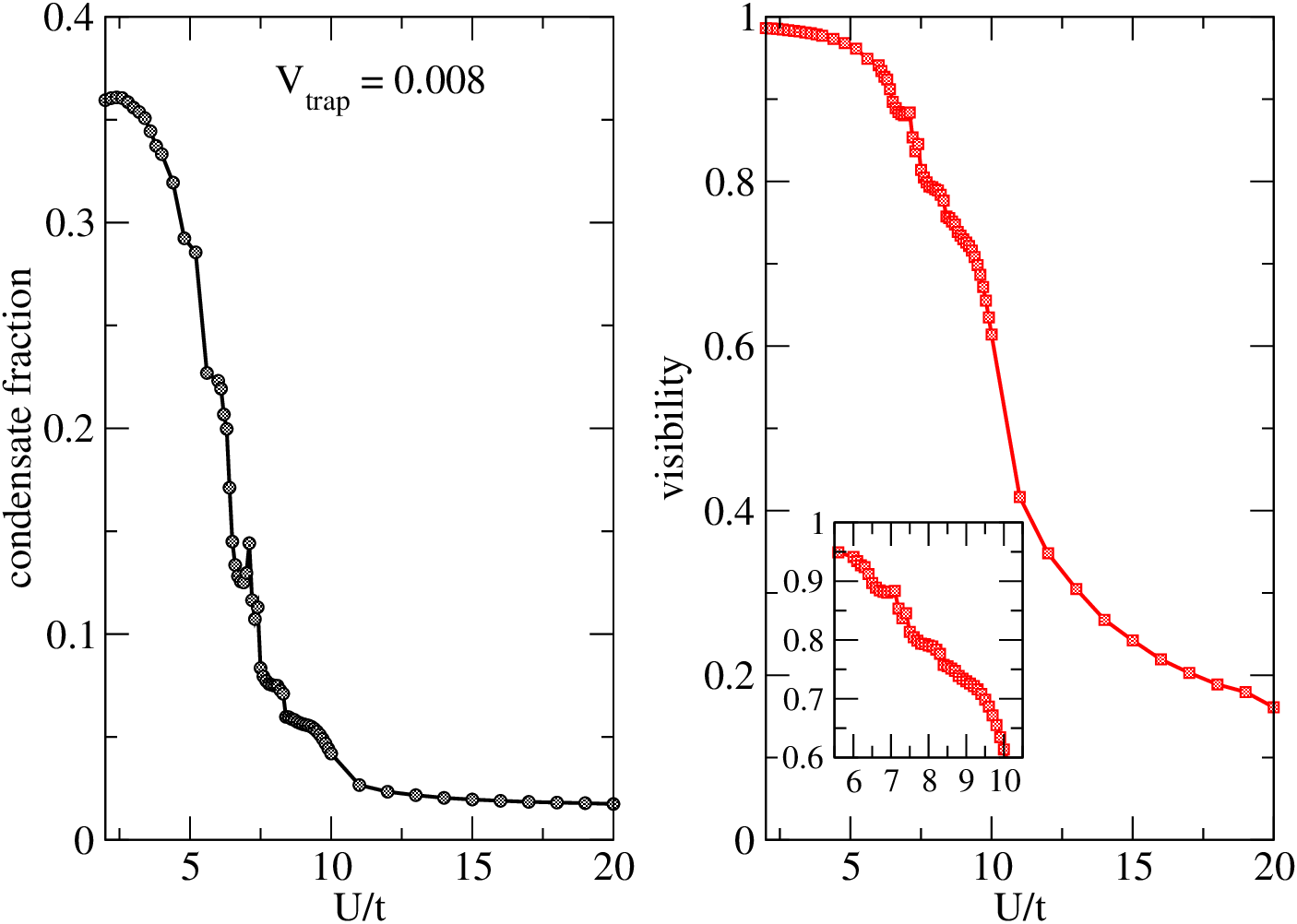}
 \caption{Condensate fraction (left) and visibility (right) as a function of $U$ for a trap of depth $\vtrap = 0.008$.}
 \label{fig:fig11}
\end{center}
\efig

\bfig[ht]
\begin{center}
 \includegraphics[width = 3.4in, angle = 0, 
clip = true]{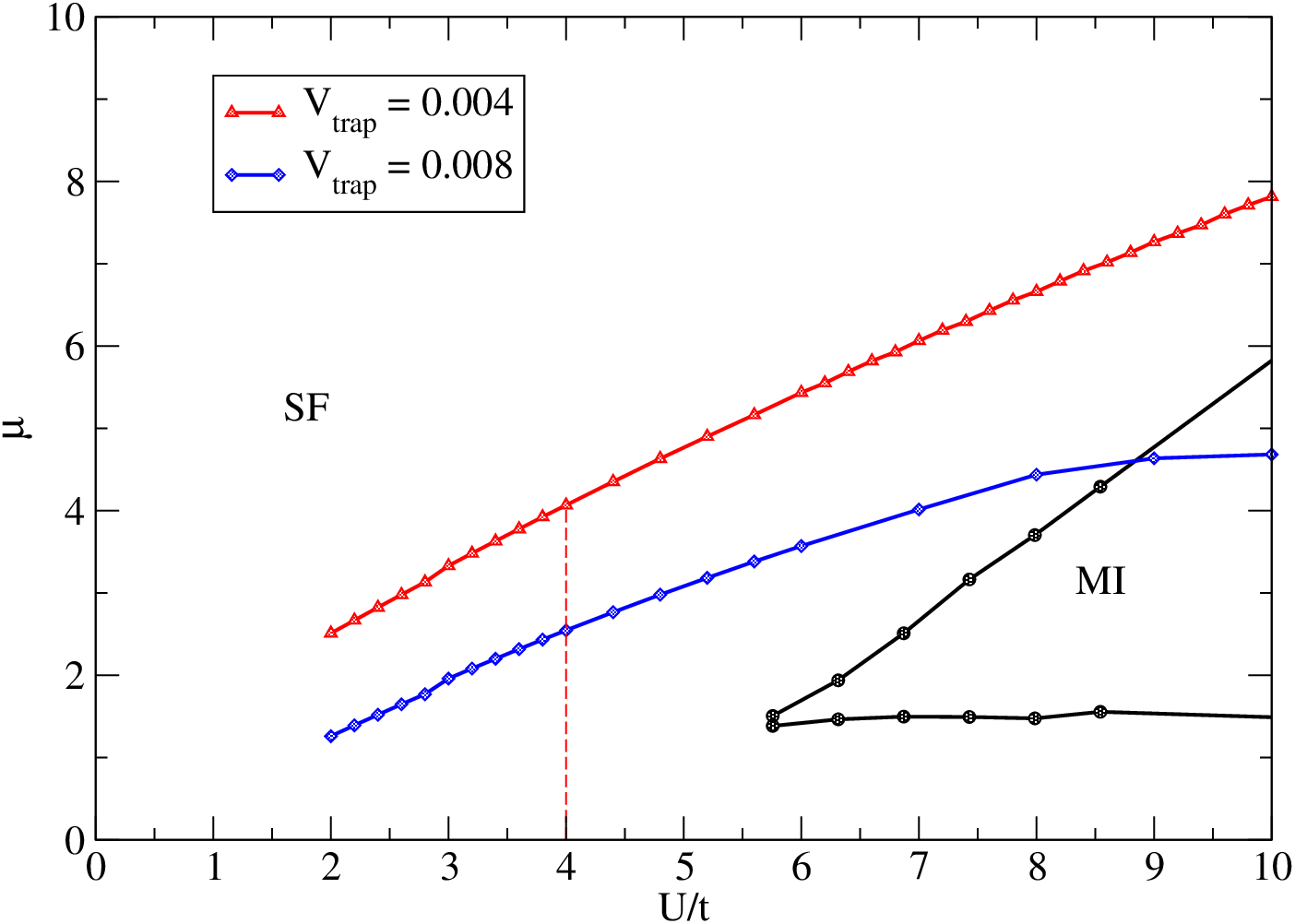}
 \caption{Variation of $\mu_0$, the 
local chemical potential at the center of the trap as a function of $U$.
For a given value of $U$, the inhomogeneous system can be represented by a vertical line originating at $\mu_0$. }
 \label{fig:fig12}
\end{center}
\efig

For a system in uniform SF phase, the fringe visibility is 
$1$~\cite{ho}. 
In the 
homogeneous case, when the system undergoes a quantum phase transition from 
SF to MI, the visibility falls monotonously~\cite{gerbier_vis, gerbier_vis2}. 
In the presence of a trap, however, the visibility as a function of $U$ has a 
rich structure due to the formation of alternating SF and MI 
shells~\cite{sengupta}. From Figs.~(\ref{fig:fig10}) and~(\ref{fig:fig11}), 
we note the following for the trap case: (1) the visibility remains finite 
even at high $U$ values (i.e., deep inside the MI lobes) compared to the 
homogeneous case, (2) kinks develop over a certain range of $U$ and (3) 
the visibility drops drastically for particular values of $U$ and for further
increase in $U$, the variation is slow. Similar behavior 
is also noted in the condensate fraction.  

For a given value of $U$ and number of bosons $N$, the homogeneous system  
represents a point ($\mu, U$) in the phase diagram given in the 
Fig.~(\ref{fig:fig5}). However, 
for an inhomogeneous system the chemical potential varies 
across the lattice and is represented by a vertical line (see Fig.~(\ref{fig:fig12})), originating at
$\mu_0$ for a given $U$ in the ($\mu-U$) plane, where
 $\mu_0$ is the chemical potential at the center of the 
trap. The values of $\mu_0$ as a function of $U$ is shown in Fig.~(\ref{fig:fig12}).
The behavior of the condensate
fraction and visibility (shown in Figs.~(\ref{fig:fig10}) and~(\ref{fig:fig11})) is then easily understood by tracing the $\mu_0$ trajectory as a function of $U$. For $U < 6.0$, a vertical line starting at $\mu_0$, representing possible values of the local chemical potential for a given $U$, does not intersect the MI lobe and no
Mott plateau forms in the density profile. As $U$ increases,
the system begins to favor the MI phase, and as a result the condensate fraction and the 
visibility decreases monotonically. However, for $U > 6.0$, the vertical line, intersects the Mott lobe, resulting in a well-developed Mott plateau in the density profile. All the kinks in the condensate
fraction and the visibility are observed for $U > 6.0$, indicating the formation and broadening of
the Mott plateau in the system, as the bosons re-distribute themselves between the two phases across the lattice. Finally for larger values of $U$,
the $\mu_0$ trajectory enters
the Mott lobe and the central 
SF region vanishes completely. The entire system is in the Mott phase except for the edges. As a result, the condensate fraction and the visibility drops drastically. Further increase in $U$ results in smooth variations of both these quantities as the SF phases exist only at the edges and the number of bosons in the 
SF phases do not vary much as shown in Fig.~(\ref{fig:fig8}).

\bfig[ht]
\begin{center}
\includegraphics[width = 3.4in, angle = 0, clip = true]{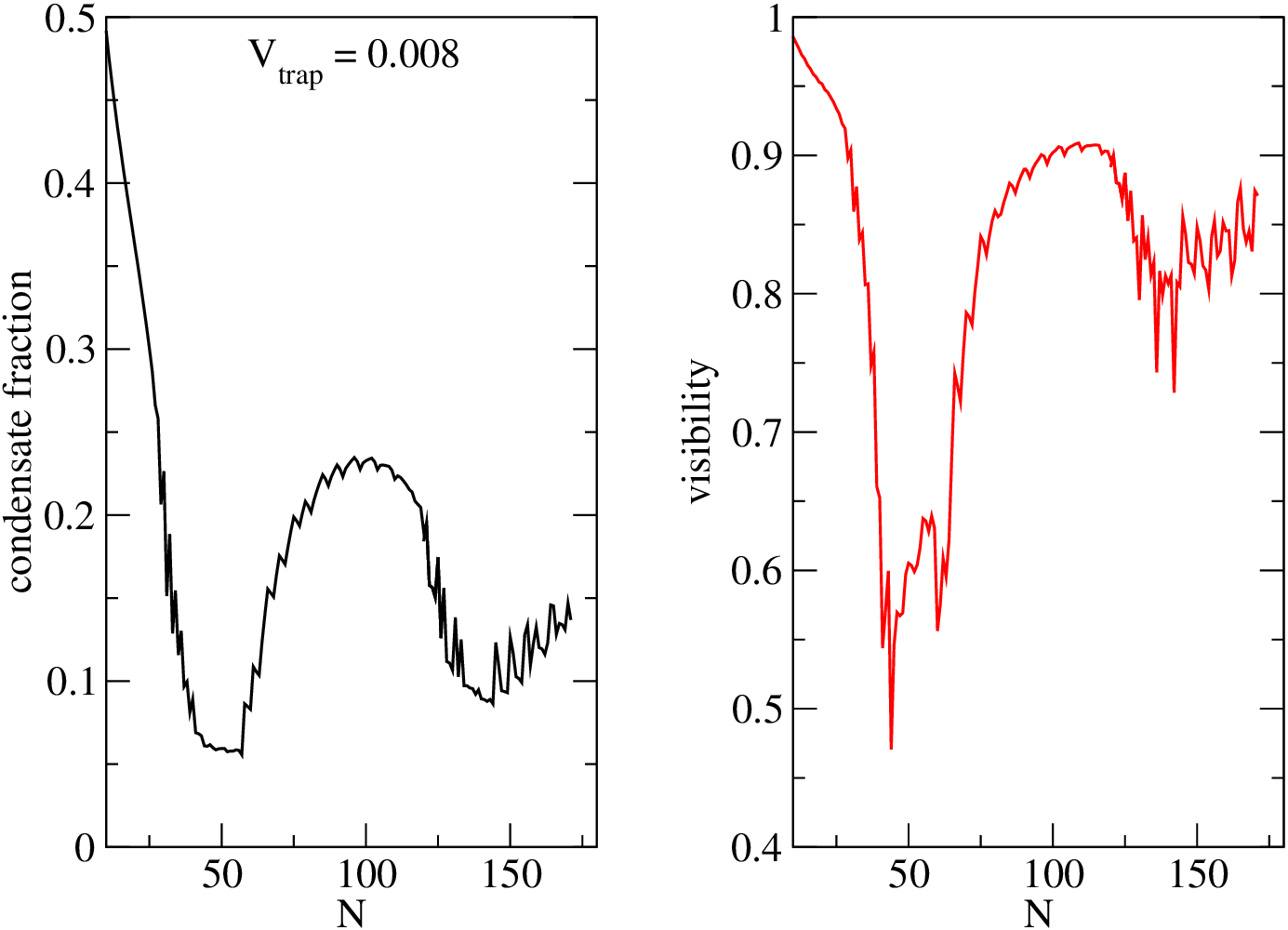}
\caption{Condensate fraction (left) and visibility (right) as a function of $N$ the total 
number of bosons in a trap of 
depth $\vtrap = 0.008$ and $U = 10$. Both the condensate 
fraction and visibility show local minima for particular ranges of $N$, indicating the formation of MI phases. 
The subsequent increase in these quantities indicate the formation of SF phases.}
\label{fig:fig13} 
\includegraphics[width = 3.4in, angle = 0, 
clip = true]{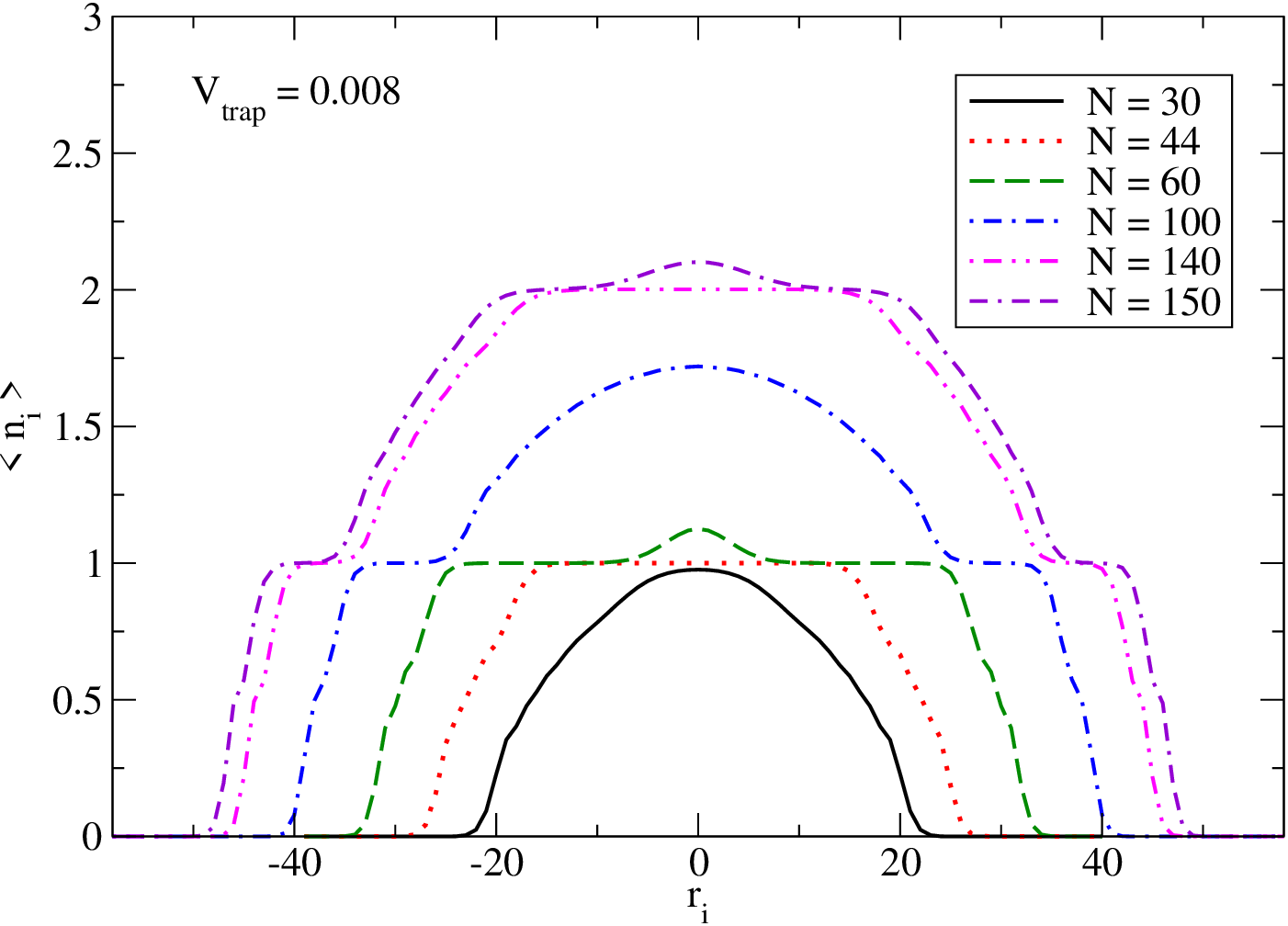}
\caption{Density profile for different total number of bosons $N$.
 Note that the 
minima in visibility and condensate fraction in Fig.~(\ref{fig:fig13}) correspond to formation of 
MI plateaus.}
\label{fig:fig14} 
\end{center}
\efig

In the experiments, the chemical potential is usually changed by changing 
the number of bosons $N$. We plot, in Fig.~(\ref{fig:fig13}), 
the variation of the condensate fraction and the visibility as a function of $N$, 
for $\vtrap = 0.008$ and $U = 10$. 
We see that when the MI plateau forms, both visibility and 
condensate fraction dip, the first of these corresponding to the formation of a $\rho = 1$ 
Mott plateau, occurs around $N \sim 40$. This can be observed in the density 
profile~(see Fig.~(\ref{fig:fig14})). The plateau in Fig.~(\ref{fig:fig13}), in the condensate fraction, indicates the range of $N$ for which the $\rho = 1$ MI phase is possible for the given $U$ value. We note that beyond
$N \sim 56$, a $1 \le \rho \le  2$ SF forms, marked by an increase in the visibility and the condensate fraction.
The second minimum 
occurs around $N \sim 140$ signals the formation of the second Mott 
plateau ($\rho = 2$), as shown in Fig.~(\ref{fig:fig14}). 
Further increase in $N$, beyond $N \sim 150$, results in 
another SF phase, with the on-site density ranging between $2 \le \rho \le3$. 
Therefore, for a fixed value of $U$, the minima in the condensate fraction and the visibility as a function 
of $N$ are good indicators for the formation of Mott plateaus. 

\section{Conclusion}
\label{sec5}
	This paper demonstrates a way of extracting the phase diagram 
for the Bose-Hubbard model in
the presence of an external trapping potential, from the number density 
profile. Signatures of these phases in experimentally observed quantities 
such as visibility, condensate fraction and $N(\rho)$ have been documented. We have also obtained 
the density-profile as a function of the chemical potential for the 
homogeneous case, using FS-DMRG for the first time, to the best of our 
knowledge. 

Future directions are immense. As a first step, the extended Bose-Hubbard 
model can be investigated and the phase diagram, together with the experimental 
signatures for the various phases can be extracted in an analogous way, 
which is in progress. The influence of a three-body term for the 
Bose-Hubbard and the extended Bose-Hubbard can be investigated, which is also 
in progress. 

The presence of a trap makes the system interesting due to the simultaneous 
existence of different phases (such as the SF and the MI phase in this work) and gets the theoretical predictions closer to experiments. It would also 
be interesting to study the Bose-Hubbard model for two boson species and 
Bose-Fermi mixtures in the presence of a trap. This paper serves more as 
benchmark to extract the phase diagram in a straight-forward and transparent 
way from the density profile. This technique can now be extended to other problems.  

\acknowledgments{The computations were carried out on the Garuda grid and we thank its members for their support. ML thanks the Indian Institute of Astrophysics, Bangalore for their hospitality, where all of this work was carried out. RVP thanks CSIR, India Grant No. 03(1107)/08/EMR-II.}



\begin{thebibliography}{99} 
\bibitem{dalfovo99} F. Dalfovo, S. Giorgini, L. P. Pitaevskii and S. Stringari, 
Rev. Mod. Phys. \textbf{71} 463 (1999).

\bibitem{block08} 
I. Bloch, J. Dalibard, and W. Zwerger, 
Rev. Mod. Phys. \textbf{80} 885 (2008). 

\bibitem{lewenstein07} M. Lewenstein, A.Sanpera, V.Ahufinger, 
B.Damski, A.Sen and U.Sen, 
Advances in Physics, \textbf{56} 243 (2007).

\bibitem{fisher} M.P.A. Fisher, P.B. Weichmann, G. Grinstein and D.S. Fisher, 
Phys. Rev. B \textbf{40}, 546 (1989).
 
 \bibitem{krauth} W.Krauth, M. Caffarel, and J. P. Bouchaud,
Phys. Rev. B \textbf{45}, 3137 (1992).

\bibitem{sheshadri}K. Sheshadri, H. R. Krishnamurthy, R. Pandit, and T. V. Ramakrishnan,
Europhys. Lett. \textbf{22}, 257 (1993).

\bibitem{amico} L. Amico and V. Penna, Phys. Rev. Lett. \textbf{80},
2189 (1998).

\bibitem{kuhner} T. D. Kuhner and H.
Monien, Phys. Rev. B \textbf{58}, R14741 (1998); T.D. Kuhner, S.R.
White, H.Monien, Phys. Rev. B. {\textbf 61},12474 (2000).

\bibitem{pandit} R. V. Pai and R. Pandit, Phys. Rev. B {\textbf 71}, 104508
(2005).

\bibitem{pai} R.V.Pai, R.Pandit, H.R. Krishnamurthy and S. Ramasesha, Phys. Rev. Lett. {\textbf 76}, 2937
(1996).

\bibitem{kashurnikov}V. A. Kashurnikov and B.
V. Svistunov, Phys. Rev. B \textbf{53}, 11776 (1996).

\bibitem{batrouni}G. G. Batrouni, R.
T. Scalettar, G. T. Zimanyi, and A. P. Kampf, Phys. Rev. Lett.
\textbf{74}, 2527 (1995); P. Niyaz, R. T. Scalettar, C. Y. Fong, and
G. G. Batrouni, Phys. Rev. B \textbf{44}, 7143 (1991).

\bibitem{freericks}J.K. Freericks and H. Monien, Europhys. Lett. \textbf{26}, 2691 (1994);
J.K. Freericks and H. Monien, Phys. Rev. B \textbf{53}, 2691 (1996).

\bibitem{qmc1}
 B. Capogrosso-Sansone, N. V. Prokofev, and B. V. Svistunov,
  Phys. Rev. B {\bf 75}, 134302 (2007).

\bibitem{qmc2} Barbara Capogrosso-Sansone, Sebnem Gunes Soyler, Nikolay Prokof'ev and Boris Svistunov, arXiv:0710.2703 [cond-mat.other]

\bibitem{jaksch}D. Jaksch, C. Bruder, J. I. Cirac, C. W. Gardiner, 
and P. Zoller,
Phys. Rev. Lett. \textbf {81}, 3108 (1998).

\bibitem{greiner} M. Greiner, O. Mandel, T. Esslinger, T. W. Haensch, 
and I. Bloch, Nature \textbf{415}, 39 (2002).

\bibitem{spielman} I. B. Spielman, W. D. Phillips, and J. V. Porto,
Phys. Rev. Lett. {\bf 98} 080404 (2007).

\bibitem{stoferle} T. Stoeferle, H. Moritz, C. Schori, M. Koehl, 
and T. Esslinger,
Phys. Rev. Lett. {\bf 92}, 130403 (2004).

\bibitem{wessel} S. Wessel, F. Alet, M. Troyer and G. G. Batrouni,
Phys. Rev. A, \textbf{70} 053615 (2004).

\bibitem{svistunov}V. A. Kashurnikov, N. V. Prokof�ev, and B. V.
Svistunov, Phys. Rev. A, \textbf{66}, 031601 (2002).

\bibitem{stoof}D. van Oosten, P. van der Straten, and H. T. Stoof, Phys. Rev. A
 \textbf{63} 053601 (2001).

\bibitem{bergkvist}S. Bergkvist, P. Henelius, and A.
Rosengren, Phy. Rev. A {\bf 70}, 053601 (2004).

\bibitem{pollet}Lode Pollet, Stefan Rombouts, Kris Heyde, and Jorge
Dukelsky, Phys. Rev. A \textbf{69}, 043601 (2004).

\bibitem{smita}B. DeMarco, C. Lannert, S. Vishveshwara, and T.-C.
Wei, Phys. Rev. A \textbf{71}, 063601 (2005).

\bibitem{mitra}Kaushik Mitra, C. J. Williams, C. A. R. S�a de Melo,
Phys. Rev. A {\bf 77}, 033607 (2008).

\bibitem{bloch06} Simon Fo�lling, Artur Widera, Torben Mu�ller, Fabrice Gerbier, and Immanuel
Bloch, Phys. Rev. Lett. \textbf{97}, 060403 (2006).

\bibitem{campbel} G. K. Campbell, J. Mun, M. Boyd, P. Medley,
A. E. Leanhardt, L. G. Marcassa, D. E. Pritchard, W. Ketterle,
Science \textbf{313}, 649 (2006).

\bibitem{gerbier_vis} Fabrice Gerbier, Artur Widera, Simon F\"{o}lling, 
Olaf Mandel, Tatjana Gericke and Immanuel Bloch, PRA \textbf{72}, 053606, (2005).

\bibitem{kt} Kosterlitz J M and Thouless D. J., J. Phys. \textbf{C6} 1181, 
(1973).

\bibitem{giamarchi} Thierry Giamarchi, {\it Quantum Physics in One Dimension}, Clarendon Press, Oxford, (2004). 

\bibitem{qmcwork}G. G. Batrouni et. al. Phys. Rev. Lett. {\bf 89}, 117203 (2002)

\bibitem{schollwock1} C. Kollath, U. Schollwöck, J. von Delft, W. Zwerger, 
Phys. Rev. A {\bf 71 69} (Rapid Communication), 031601 (2004).  

\bibitem{epl}Laura Urba {\it et al}
J. Phys. B: At. Mol. Opt. Phys. {\bf 39} 5187 (2006).  

\bibitem{hrk} G.G. Batrouni, H. R. Krishnamurthy, K. W. Mahmud, 
V.G. Rousseau, R.T. Scalettar Phy. Rev. A {\bf 78}, 023627 (2008).


\bibitem{white} S. R. White, Phys. Rev. Lett. {\bf 69}, 2863 (1992); 
Phys. Rev. B {\bf 48},  10345 (1993).

\bibitem{schollwock} U. Schollw\"{o}ck, Rev. Mod. Phys. {\bf 77}, 259 (2005).


\bibitem{sengupta} P.\ Sengupta, M.\ Rigol, G.\ G.\ Batrouni,
 P.\ J.\ H.\ Denteneer and R.\ T.\ Scalettar, Phys. Rev. Letts. 
\textbf{95}, 220402, (2005).

\bibitem{ho} R. B.\ Diener, Q. Zhou, H. Zhai and Tin-Lin Ho, Phys. Rev. Lett. \textbf{98}, 180404 (2007) arXiv:cond-mat/0609685

\bibitem{nandini} Y. Kato, Q. Zhou, N. Kawashima and N. Trivedi, 
Nature Physics  \textbf{4}, 617, (2008). 

\bibitem{gerbier_vis2} F. Gerbier, A. Widera, S. F\"{o}lling, O. Mandel, 
T. Gericke and I. Bloch, Phys. Rev. Letts. \textbf{95}, 050404 (2005). 

\bibitem{leggett_rev} A.\ J.\ Leggett, Rev.\ of Mod.\ Phys., \textbf{73} (2001).

\end{thebibliography}
\end{document}